\documentclass[final,3p,authoryear, times]{elsarticle}
\usepackage{color}
\usepackage{multirow,setspace,times,amssymb,amsmath,graphicx,color,subfigure,url}
\usepackage{dsfont}
\usepackage{threeparttable}
\usepackage{rotating}
\usepackage{lscape,pdflscape}
\usepackage{dcolumn}
\usepackage{listings}
\usepackage{relsize}
\usepackage[colorlinks=true,linkcolor=blue,citecolor=blue,urlcolor=blue]{hyperref}
\usepackage{tikz}

\journal{Elsevier}

\begin{document}

\begin{frontmatter}

\title{Visibility graph analysis of crude oil futures markets: Insights from the COVID-19 pandemic and Russia-Ukraine conflict}

\author[SILC]{Yan-Hong Yang}
\author[SFT]{Ying-Lin Liu\corref{cor1}}
\ead{yinglinl@outlook.com}
\author[SUIBE]{Ying-Hui Shao \corref{cor1}}
\ead{yinghuishao@126.com}

\cortext[cor1]{Corresponding author. }

\address[SILC]{SILC Business School, Shanghai University, Shanghai 201899, China}
\address[SFT]{School of Financial Technology, Shanghai Lixin University of Accounting and Finance, Shanghai 201209,China}
\address[SUIBE]{School of Statistics and Information, Shanghai University of International Business and Economics, Shanghai 201620, China}

\begin{abstract}
	
Drawing inspiration from the significant impact of the ongoing Russia-Ukraine conflict and the recent COVID-19 pandemic on global financial markets, this study conducts a thorough analysis of three key crude oil futures markets: WTI, Brent, and Shanghai (SC). Employing the visibility graph (VG) methodology, we examine both static and dynamic characteristics using daily and high-frequency data. We identified a clear power-law decay in most VG degree distributions and highlighted the pronounced clustering tendencies within crude oil futures VGs. Our results also confirm an inverse correlation between clustering coefficient and node degree and further reveal that all VGs not only adhere to the small-world property but also exhibit intricate assortative mixing. Through the time-varying characteristics of VGs, we found that WTI and Brent demonstrate aligned behavior, while the SC market, with its unique trading mechanics, deviates. The 5-minute VGs' assortativity coefficient provides a deeper understanding of these markets' reactions to the pandemic and geopolitical events. Furthermore, the differential responses during the COVID-19 and Russia-Ukraine conflict underline the unique sensitivities of each market to global disruptions. Overall, this research offers profound insights into the structure, dynamics, and adaptability of these essential commodities markets in the face of worldwide challenges.

\end{abstract}
\end{frontmatter}

\section{Introduction}
\label{S1:Introduction}

In recent years, the COVID-19 pandemic and the Russia-Ukraine conflict have stood out as monumental global events, sending ripples across international financial landscapes, especially within the realm of crude oil futures. Crude oil, a pivotal energy source, is integral to a nation's economic trajectory. Its price volatility can expose participants in foreign exchange, stock, and other financial markets to significant risks \citep{he2021correlation,shao2021new,zhu2021multidimensional, shao2021Fractals,ji2020searching}. 

Notably, the onset of the COVID-19 pandemic precipitated a drastic drop in crude oil prices. In a shocking turn of events on April 20, 2020, the contract price for West Texas Intermediate (WTI) crude oil plunged to an unprecedented -\$37.63 per barrel, marking the first time in history it ventured below zero.
The unprecedented plunge in oil prices due to the pandemic not only made headlines but also became a focal point of academic research. The impact of the COVID-19 crisis on crude oil has raised wide attention from academia. \cite{narayan2020oil} assessed the effect of COVID-19 infections and oil price news in influencing oil prices. They unveil that the COVID-19 pandemic and negative oil price news will affect oil price under a higher oil price volatility. \cite{gil2020crude} analyzed the shock of COVID-19 on crude oil prices using long memory approaches. They conclude that oil price series is mean everting and the impact of COVID-19 will be transitory. Besides, \cite{gharib2021impact} analyzed explosive behavior and bubbles in oil price by using LPPLS model. The findings suggest that the WTI and Brent crude oil prices are significantly driven by bubbles during the COVID-19.
In addition, there exist many studies on the risk spillover nexus between crude oil and stock markets during the COVID-19. For instance, \cite{zhu2021multidimensional} have studied multidimensional risk spillovers between crude oil and stock markets during the COVID-19 pandemic. They reveal significant spillovers from stock markets to oil markets. And their findings indicate that the oil-stock risk spillovers during the COVID-19 are obviously stronger than those in normal times. Similarly, \cite{zhang2021crude} analyzed the return and volatility spillover among crude oil market, stock market and the COVID-19 pandemic across the United States, Japan and Germany. They present that the influence of COVID-19 pandemic on the stock and oil markets exceeds that of the subprime crisis of 2008. \cite{mensi2021volatility} investigated frequency spillovers and connectedness between oil and stock markets. They found the COVID-19 pandemic has intensified total spillovers among the markets.

Nonetheless, the studies mentioned above predominantly address the effects of COVID-19 on developed crude oil markets, notably WTI and Brent. On March 26, 2018, China inaugurated the Shanghai (SC) crude oil futures, marking the advent of an emerging commodity futures market. Consequently, an abundance of studies has also centered on the impact of COVID-19 on the SC crude oil futures market. \cite{yang2021extreme} explored the risk spillovers between SC crude oil futures and international oil futures, discerning that SC oil futures play a role as a net risk recipient within the global oil system. Concurrently, in the wake of the COVID-19 outbreak, the risk spillover between SC crude oil futures and international oil futures has seen a marked escalation since the onset of 2020. \cite{chen2022forecasting} delved into the role of jumps and leverage in forecasting the realized volatility of China's crude oil futures, determining that forecasting models incorporating leverage effects exhibited optimal predictive capabilities during the COVID-19 pandemic. Research by \cite{hu2023time} noted an augmentation in the jump intensity and magnitude of the Shanghai crude oil futures market due to COVID-19. Drawing on the multifractal structural changes in the SC market, \cite{shao2023short} identified a significant enhancement in the market efficiency of SC and its interrelations with other assets post the outbreak of COVID-19. \cite{zhang2022impact} probed the impact of COVID-19 on the interdependence between Chinese and U.S. oil futures markets, and their findings underscored that the COVID-19 pandemic bolstered the long-term correlation between the two oil markets.

Concurrently, the Russia-Ukraine conflict, while not intrinsically energy-centric, escalated geopolitical tensions on an international scale. Given Ukraine's strategic positioning and its quintessential role in the European energy supply chain, this confrontation indirectly resonated within the crude oil markets. Specifically, sanctions against Russia and their potential ramifications instilled apprehensions about future energy supplies, engendering price volatility. That is, the global geopolitical uncertainty, with the Russia-Ukraine conflict as its primary catalyst, has further exacerbated the volatility within the crude oil markets, hindering the global economic recovery post-COVID-19. This has also sparked a subsequent discourse within financial literature \citep{boubaker2022heterogeneous,zhang2023impact,inacio2023assessing}. \cite{pan2023changes} investigated the alterations in volatility leverage and spillover effects of crude oil futures markets influenced by the Russia-Ukraine conflict. Their research revealed that this geopolitical strife significantly altered the leverage effects across these markets and marginally increased and stabilized the dynamic conditional correlations among them, while diminishing the volatility spill-over effects. \cite{cui2023higher} examined the higher-order moment risk connectedness among WTI futures, Brent futures, SC futures, and other commodity futures both pre- and post-COVID-19, as well as following the Russia-Ukraine conflict's onset. Their empirical findings suggest a positive, time-varying dynamic linkage between international oil and major commodity futures, which was considerably strengthened amidst the COVID-19 pandemic and the Russia-Ukraine crisis. \cite{huang2023correlations}, utilizing the multifractal method, analyzed the repercussions of the Russia-Ukraine conflict on the crude oil market and the contagion effects on the stock markets of importing and exporting nations. Their study indicates that the efficiency of the crude oil market post-conflict was inferior to its pre-conflict state. Moreover, they observed a heightened interrelationship between the crude oil market and the stock market for oil-importing countries post-conflict, while no significant shift was discerned in the mutual relationships between the crude oil market and stock market for oil-exporting nations.

When the demand shocks of the pandemic dovetailed with the supply risks emanating from the Russia-Ukraine skirmish, the crude oil futures market witnessed pronounced turbulence. This climate of uncertainty and volatility posed formidable challenges for investors, policymakers, and corporations alike. Yet, it also proffered an invaluable opportunity for researchers to discern how global financial markets navigate the waters of multifaceted shocks.

In this study, our primary contribution is to explore the effects of crisis events on crude oil futures markets through the lens of network topology evolution. This offers a fresh perspective for risk mitigation in the face of significant crisis events. It's worth noting that the visibility graph method we employed was originally introduced by \cite{lacasa2008time}, serving as an innovative approach to time series analysis within the realm of complex networks.
Visibility graph (VG) is especially advantageous when applied to crude oil futures markets and other financial sectors \citep{dai2019visibility,liu2020visibility,sun2016visibility}. They adeptly identify complex relationships amidst price variations, provide a visual representation of market behavior, and enable the analysis of temporal dynamics. These graphs are particularly useful for handling high-frequency data and facilitate in-depth exploration of critical network attributes, contributing to a comprehensive understanding of market behavior. In summation, visibility graphs are a powerful tool for gaining insights into the complex and dynamic nature of crude oil futures markets.

Overall, the crude oil futures market has experienced notable disruptions from the COVID-19 pandemic and the Russia-Ukraine conflict, affecting price trajectories and market participants' expectations. We highlight our contributions as follows: Using the visibility graph methodology, we transform crude oil futures prices into spatial networks, highlighting dynamics in both emerging (SC) and established markets (WTI, Brent). To our knowledge, this is the inaugural study analyzing the Shanghai crude oil market through a visibility graph lens. We delve into VGs' static and evolving attributes across varied data resolutions and assess the distinct impacts of the COVID-19 crisis and the Russia-Ukraine conflict on the three oil markets. Our analysis provides a comparative view of the market repercussions of these pivotal events.

 The rest of this paper is organized as follows. Section~\ref{S2:Data description} depicts the data sets and presents the summary statistics. Section~\ref{S3:Methodology} explains the methodology. Section~\ref{S4:Empirical analysis} reports the empirical results, and Section~\ref{S5:conclusion} concludes with a review of the main findings.

\section{Data description and summary statistics}
\label{S2:Data description}

Throughout this research, we select WTI, Brent, and SC as the three benchmark indices for global crude oil futures. Our data was sourced from the Wind database. To reveal the underlying influence mechanisms of the visibility graph network across different crude oil futures markets comprehensively, we use research samples at various frequencies. Specifically, our sample data comprises price series at daily, 5-minute, 15-minute, and 30-minute intervals. Consistent with many studies, we choose the 5-minute, 15-minute, and 30-minute frequencies to minimize noise \citep{shao2023short}. We use the front-month futures contracts to form continuous closing price time series for all three assets. Our sample covers the period from March 26, 2018, the official launch date of China's yuan-denominated crude oil futures, to July 20, 2023, spanning two major crises: the COVID-19 pandemic and the Russia-Ukraine conflict. It should be noted that due to data availability, there are two gaps in the 5-minute, 15-minute, and 30-minute closing prices for the SC crude oil futures: from March 26, 2018, to September 3, 2019, and from July 3, 2021, to December 5, 2021.

 \begin{figure}[!htb]
	\centering
	\includegraphics[width=14cm]{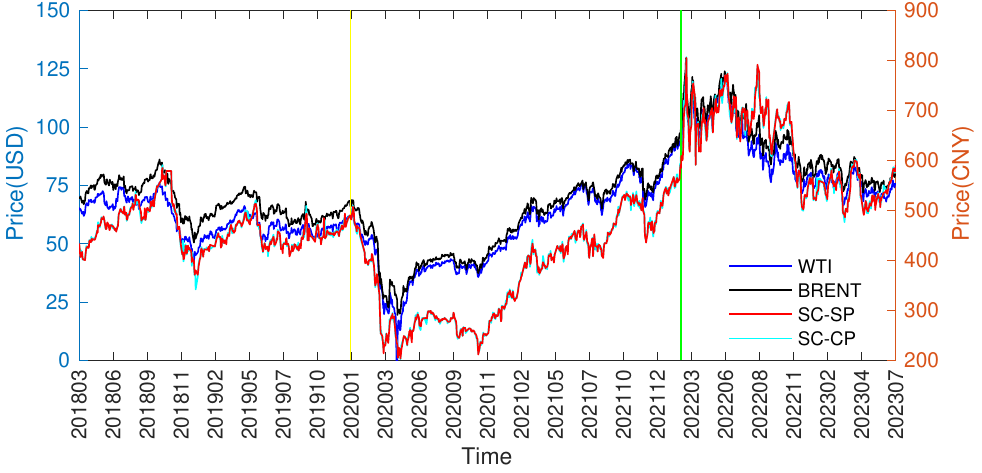}
	\caption{Time series of daily prices for the three crude oil futures markets over the sample period: March 26, 2018, to July 20, 2023.}
	\label{Fig:VG:SC:WTI:BRENT:Price}
\end{figure}

Considering the impact of the COVID-19 pandemic and the Russia-Ukraine conflict on these three crude oil markets, we have divided the daily sample data into four sub-samples: namely, Sub-sample 1, Sub-sample 2, Sub-

\begin{landscape}
	\begin{table}[!htp]
	\footnotesize
	\relsize{-1.5}
	\caption{Descriptive statistics and preliminary tests on these three crude oil future price series.}
	\label{TB:VG:Summary:Statistics}
		\medskip
		\centering
		\begin{threeparttable}
			\begin{tabular}{llllllllllllllllllllllllllllllllllllll}
				\hline\hline
				\multirow{3}*[2mm]{Name}&\multicolumn{5}{l}{SC-SP}&&\multicolumn{5}{l}{SC-CP}&&\multicolumn{5}{l}{WTI}&&\multicolumn{5}{l}{Brent}\\
				\cline{2-6}\cline{8-12}\cline{14-18}\cline{20-24}
				& Sub1 & Sub2 & Sub3 & Sub4 & Whole && Sub1 & Sub2 & Sub3 & Sub4 & Whole && Sub1 & Sub2 & Sub3 & Sub4 & Whole && Sub1 & Sub2 & Sub3 & Sub4 & Whole  \\
				\hline
				\multicolumn{16}{l}{\textit{Panel A : Daily data}} \\
				Observations & 432& 519& 210 & 132 & 1293 && 432& 519& 210 & 132 & 1293&& 450& 556& 223 & 144 & 1373&& 456& 557& 220 & 142 & 1375\\
				Mean & 462.0& 375.5& 670.6 & 542.1 & 469.3 && 462.1& 375.6& 669.9 & 542.5 & 469.3&& 60.8& 56.2& 95.8 & 74.7 & 66.1&& 68.0& 59.3& 100.8 & 79.8 & 71.0\\
				Maximum & 584.8& 590.3& 805.0 & 600.4 & 805.0 && 590.6& 593.5& 806.6 & 601.3 & 806.6&& 76.2& 94.8& 124.8 & 83.3 & 124.8&& 86.1& 97.5& 129.5 & 88.2 & 129.5\\
				Minimum & 371.2& 203.9& 502.2 & 486.3 & 203.9 && 342.0& 202.3& 499.3 & 476.6 & 202.3&& 42.7& -13.1& 71.6 & 66.3 & -13.1&& 50.5& 19.5& 76.5 & 71.9 & 19.5\\
				Std.Dev. & 39.85& 99.41& 62.54 & 28.18 & 126.02 && 39.98& 99.60& 63.00 & 28.65 & 125.95&& 7.05& 18.56& 12.87 & 4.10 & 19.59&& 7.19& 17.96& 11.82 & 4.42 & 19.50\\
				Skewness & 0.767& 0.106& -0.785 & 0.137 & 0.166 && 0.722& 0.110& -0.778 & 0.101 & 0.161&& 0.125& -0.191& 0.170 & 0.074 & 0.200&& 0.240& -0.098& 0.059 & 0.125 & 0.113\\
				Kurtosis & 3.769& 1.723& 3.314 & 2.146 & 2.793 && 3.947& 1.726& 3.315 & 2.170 & 2.792&& 2.096& 2.478& 2.038 & 1.976 & 3.487&& 2.132& 2.049& 2.201 & 1.737 & 3.247\\
				$p$-$\rm value_{\rm JB}$ & 0.000& 0.000& 0.002 & 0.077 & 0.019 && 0.000& 0.000& 0.002 & 0.092 & 0.021&& 0.003& 0.013& 0.016 & 0.038 & 0.000&& 0.002& 0.000& 0.045 & 0.016 & 0.040\\
				$p$-$\rm value_{\rm ADF}$ & 0.191& 0.837& 0.905 & 0.231 & 0.575 && 0.138& 0.841& 0.888 & 0.172 & 0.567&& 0.362& 0.890& 0.669 & 0.104 & 0.446&& 0.364& 0.966& 0.893 & 0.200 & 0.527\\
				\\
				\multicolumn{4}{l}{\textit{Panel B: 5-minute data}} \\\
				\multirow{3}*[2mm]{}&&\multicolumn{6}{l}{SC}&&\multicolumn{6}{l}{WTI}&&\multicolumn{6}{l}{Brent}\\
				\cline{2-7}\cline{9-14}\cline{16-21}
				& M1 & M2 & M3 & M4 & M5& Whole && M1 & M2 & M3 & M4 & M5&  Whole && M1 & M2 & M3 & M4 & M5&  Whole \\
				\hline
				Observations & 1713& 1114& 1493 & 1887 & 1066 & 63034&& 5565& 6024 & 6042 & 5988& 5574& 378723&& 5246 & 5751 & 5819& 5743& 5283& 361046 \\
				Mean & 470.5& 478.3& 553.5 & 690.0 & 543.8 & 480.7&& 59.6& 58.2 & 89.9 & 107.2& 78.5& 66.0&& 64.4 & 63.9 & 91.2& 109.4& 84.1& 70.7 \\
				Maximum & 499.9& 516.3& 593.5 & 822.4 & 569.7 & 822.4&& 62.3& 65.5 & 97.2 & 129.2& 82.5& 129.2&& 67.5 & 70.9 & 99.2& 132.8& 88.9& 132.8 \\
				Minimum & 445.8& 453.0& 523.8 & 584.2 & 513.8 & 199.4&& 55.4& 51.7 & 83.1 & 90.1& 72.6& 6.9&& 60.4 & 56.8 & 85.3& 92.8& 77.8& 16.0 \\
				Std.Dev. & 13.59& 16.84& 13.08 & 55.66 & 16.25 & 152.48&& 1.67& 3.36 & 2.47 & 8.85& 2.72& 19.34&& 1.64 & 3.41 & 2.41& 9.49& 3.14& 19.06\\
				Skewness & -0.056& 0.631& 0.331 & 0.410 & -0.235 & 0.018&& -0.683& -0.147 & -0.042 & 0.202& -0.575& 0.242&& -0.476 & -0.294 & 0.136& 0.409& -0.537& 0.084\\
				Kurtosis & 1.711& 2.115& 3.305 & 2.397 & 1.593 & 1.926&& 2.696& 2.166 & 2.701 & 2.377& 1.959& 3.353&& 2.803 & 2.242 & 2.468& 2.453& 1.894& 3.174\\
				$p$-$\rm value_{\rm JB}$ & 0.000& 0.000& 0.000 & 0.000 & 0.000 & 0.000&& 0.000& 0.000 & 0.000 & 0.000& 0.000& 0.000&& 0.000 & 0.000 & 0.000& 0.000& 0.000& 0.000\\
				$p$-$\rm value_{\rm ADF}$ & 0.929& 0.728& 0.408 & 0.440 & 0.498 & 0.635&& 0.304& 0.907 & 0.362 & 0.426& 0.511& 0.433&& 0.427 & 0.926 & 0.584& 0.465& 0.604& 0.418\\
				\\
				\multicolumn{4}{l}{\textit{Panel C: 15-minute data}} \\\
				\multirow{3}*[2mm]{}&&\multicolumn{6}{l}{SC}&&\multicolumn{6}{l}{WTI}&&\multicolumn{6}{l}{Brent}\\
				\cline{2-7}\cline{9-14}\cline{16-21}
				& M1 & M2 & M3 & M4 & M5& Whole && M1 & M2 & M3 & M4 & M5&  Whole && M1 & M2 & M3 & M4 & M5&  Whole \\
				\hline
				Observations & 614& 425& 533 & 675 & 393 & 22797&& 1855& 2008 & 2014 & 1996& 1858& 126248&& 1786 & 1948 & 1960& 1938& 1795& 122153 \\
				Mean & 470.3& 478.2& 554.7 & 691.1 & 542.2 & 480.7&& 59.6& 58.2 & 89.9 & 107.2& 78.5& 66.0&& 64.4 & 63.9 & 91.2& 109.4& 84.1& 70.7 \\
				Maximum & 499.0& 516.1& 593.5 & 821.7 & 567.1 & 821.7&& 62.2& 65.5 & 97.2 & 128.9& 82.5& 128.9&& 67.5 & 70.8 & 99.2& 132.7& 88.8& 132.7 \\
				Minimum & 445.9& 453.0& 523.8 & 585.0 & 515.8 & 199.4&& 55.5& 51.7 & 83.2 & 90.4& 72.8& 9.4&& 60.4 & 56.8 & 85.5& 93.0& 77.8& 16.0 \\
				Std.Dev. & 13.18& 16.73& 13.92 & 56.39 & 16.33 & 152.88&& 1.67& 3.36 & 2.47 & 8.86& 2.72& 19.31&& 1.63 & 3.40 & 2.41& 9.49& 3.14& 19.05\\
				Skewness & -0.006& 0.599& 0.403 & 0.305 & -0.042 & 0.017&& -0.674& -0.150 & -0.033 & 0.204& -0.573& 0.249&& -0.480 & -0.293 & 0.149& 0.413& -0.544& 0.086\\
				Kurtosis & 1.811& 2.089& 3.253 & 2.277 & 1.508 & 1.927&& 2.691& 2.167 & 2.715 & 2.380& 1.957& 3.346&& 2.829 & 2.247 & 2.494& 2.464& 1.906& 3.177\\
				$p$-$\rm value_{\rm JB}$ & 0.000& 0.000& 0.003 & 0.000 & 0.000 & 0.000&& 0.000& 0.000 & 0.029 & 0.000& 0.000& 0.000&& 0.000 & 0.000 & 0.000& 0.000& 0.000& 0.000\\
				$p$-$\rm value_{\rm ADF}$ & 0.939& 0.777& 0.422 & 0.306 & 0.442 & 0.663&& 0.317& 0.907 & 0.366 & 0.414& 0.511& 0.429&& 0.438 & 0.928 & 0.595& 0.463& 0.609& 0.415\\
				\\
				\multicolumn{4}{l}{\textit{Panel D: 30-minute data}} \\\
				\multirow{3}*[2mm]{}&&\multicolumn{6}{l}{SC}&&\multicolumn{6}{l}{WTI}&&\multicolumn{6}{l}{Brent}\\
				\cline{2-7}\cline{9-14}\cline{16-21}
				& M1 & M2 & M3 & M4 & M5& Whole && M1 & M2 & M3 & M4 & M5&  Whole && M1 & M2 & M3 & M4 & M5&  Whole \\
				\hline
				Observations & 330& 228& 289 & 367 & 217 & 12358&& 928& 1004 & 1007 & 998& 929& 63131&& 900 & 976 & 983& 972& 905& 61378 \\
				Mean & 470.3& 477.8& 555.4 & 691.7 & 541.4 & 481.1&& 59.6& 58.2 & 89.9 & 107.2& 78.5& 66.0&& 64.4 & 63.9 & 91.2& 109.4& 84.1& 70.7 \\
				Maximum & 499.0& 516.1& 593.5 & 820.9 & 567.1 & 820.9&& 62.2& 65.5 & 97.2 & 128.9& 82.5& 128.9&& 67.5 & 70.8 & 99.0& 132.2& 88.8& 132.2 \\
				Minimum & 445.9& 453.0& 524.9 & 585.0 & 515.8 & 202.2&& 55.5& 51.7 & 83.3 & 90.6& 72.9& 9.4&& 60.4 & 56.8 & 85.6& 93.1& 77.9& 16.0 \\
				Std.Dev. & 12.90& 16.73& 14.31 & 56.98 & 16.34 & 152.85&& 1.67& 3.37 & 2.47 & 8.87& 2.72& 19.31&& 1.63 & 3.40 & 2.42& 9.49& 3.15& 19.05\\
				Skewness & 0.020& 0.588& 0.409 & 0.245 & 0.053 & 0.016&& -0.671& -0.149 & -0.019 & 0.202& -0.572& 0.249&& -0.480 & -0.294 & 0.162& 0.410& -0.544& 0.086\\
				Kurtosis & 1.905& 2.101& 3.159 & 2.202 & 1.511 & 1.929&& 2.685& 2.167 & 2.724 & 2.381& 1.959& 3.347&& 2.836 & 2.250 & 2.506& 2.465& 1.909& 3.177\\
				$p$-$\rm value_{\rm JB}$ & 0.003& 0.002& 0.021 & 0.006 & 0.002 & 0.000&& 0.000& 0.000 & 0.183 & 0.000& 0.000& 0.000&& 0.000 & 0.000 & 0.003& 0.000& 0.000& 0.000\\
				$p$-$\rm value_{\rm ADF}$ & 0.923& 0.774& 0.446 & 0.120 & 0.634 & 0.648&& 0.385& 0.878 & 0.340 & 0.412& 0.525& 0.426&& 0.483 & 0.925 & 0.578& 0.473& 0.626& 0.429\\
				\hline\hline
			\end{tabular}
			\begin{tablenotes}[para,flushleft]
				Notes: $p$-$\rm value_{\rm JB}$ is the $p$-value of the Jarque-Bera normality test. $p$-$\rm value_{\rm ADF}$ is the $p$-value corresponding to the Augmented Dickey-Fuller unit root test.  \
			\end{tablenotes}
	\end{threeparttable}
	\end{table}
\end{landscape}

\noindent
sample 3, and Sub-sample 4. Given that the outbreak of the COVID-19 pandemic occurred on December 31, 2019, and the Russia-Ukraine conflict began on February 24, 2022, the four sub-samples are sequentially defined as follows: Sub1 ranges from March 26, 2018, to December 30, 2019; Sub2 ranges from December 31, 2019, to February 23, 2022; Sub3 spans from February 24, 2022, to December 31, 2022; and Sub4 covers January 1, 2023, to July 20, 2023.

In addition, we utilize high-frequency minute sample data to characterize the short-term effects of the COVID-19 pandemic and the Russia-Ukraine conflict on WTI, Brent, and SC within a monthly window. Given the timing and catastrophic effects of these two major crisis events, we have selected five representative sub-samples: three representing the market's normal phase and two representing the crisis event phases, to study their short-term influence. For convenience, these five selected monthly sub-samples are abbreviated as M1, M2, M3, M4, and M5, respectively. M1 covers the period from December 1, 2019, to December 30, 2019; M2 ranges from December 31, 2019, to January 30, 2020; M3 corresponds to the period from January 25, 2022, to February 23, 2022; M4 spans from February 24, 2022, to March 25, 2022; and M5 lasts from January 1, 2023, to January 31, 2023.

Fig.\ref{Fig:VG:SC:WTI:BRENT:Price} displays the daily price evolution of three crude oil futures from March 26, 2018, to July 20, 2023. Due to different settlement currencies of the three crude oil futures—with WTI and Brent priced in US dollars and SC in RMB—we place the US dollar scale on the left and the RMB scale on the right side of the vertical axis. Besides, the two vertical yellow and green lines in Fig.\ref{Fig:VG:SC:WTI:BRENT:Price} represent the onset of the COVID-19 pandemic and the Russia-Ukraine conflict, respectively, dividing the entire daily sample into four sub-samples. To highlight the VG characteristics of the newly listed crude oil future, SC, its daily settlement price series is included for comparison in the subsequent analysis. It's evident that crude oil futures faced a steep decline following the advent of COVID-19. WTI even plummeted into negative territory on April 20, 2021, registering a record settlement price of -\$37.63. After enduring the harshest months of the COVID-19 crisis, Fig.\ref{Fig:VG:SC:WTI:BRENT:Price} shows that starting from November 2020, crude oil futures prices began to oscillate and rise. However, with the outbreak of the Russia-Ukraine conflict, the prices experienced pronounced volatility before beginning a downward trend. Hence, Fig.~\ref{Fig:VG:SC:WTI:BRENT:Price} underscores the profound influence of these two major crises on the crude oil futures market.

Table~\ref{TB:VG:Summary:Statistics} presents the descriptive statistics and preliminary tests for the three crude oil future price series. At a 5\% significance level, all price sample data sets reject the assumption of a normal distribution, as evidenced by the Jarque-Bera test, as well as the skewness and kurtosis values of the data. Furthermore, the Augmented Dickey-Fuller (ADF) tests indicate that all series are nonstationary.

\begin{figure}[!htp]
	\centering
	\includegraphics[width=8cm]{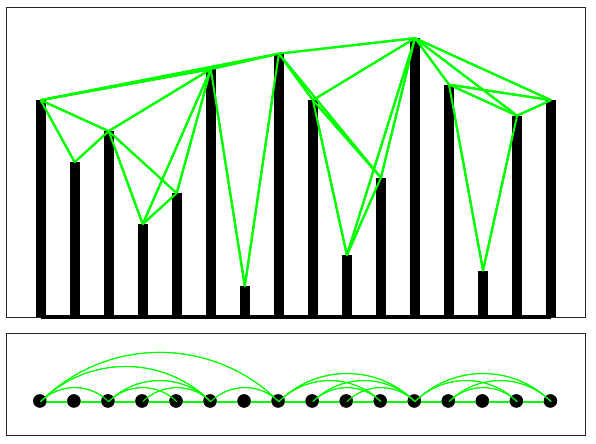}
	\caption{Schematic representation of the visibility algorithm. The upper panel corresponds to a time series containing 16 data points, and the associated visibility graph is presented in lower panel.}
	\label{Fig:VG:Schematic}
\end{figure}

 \begin{figure}[!htp]
	\centering
	\includegraphics[width=12cm]{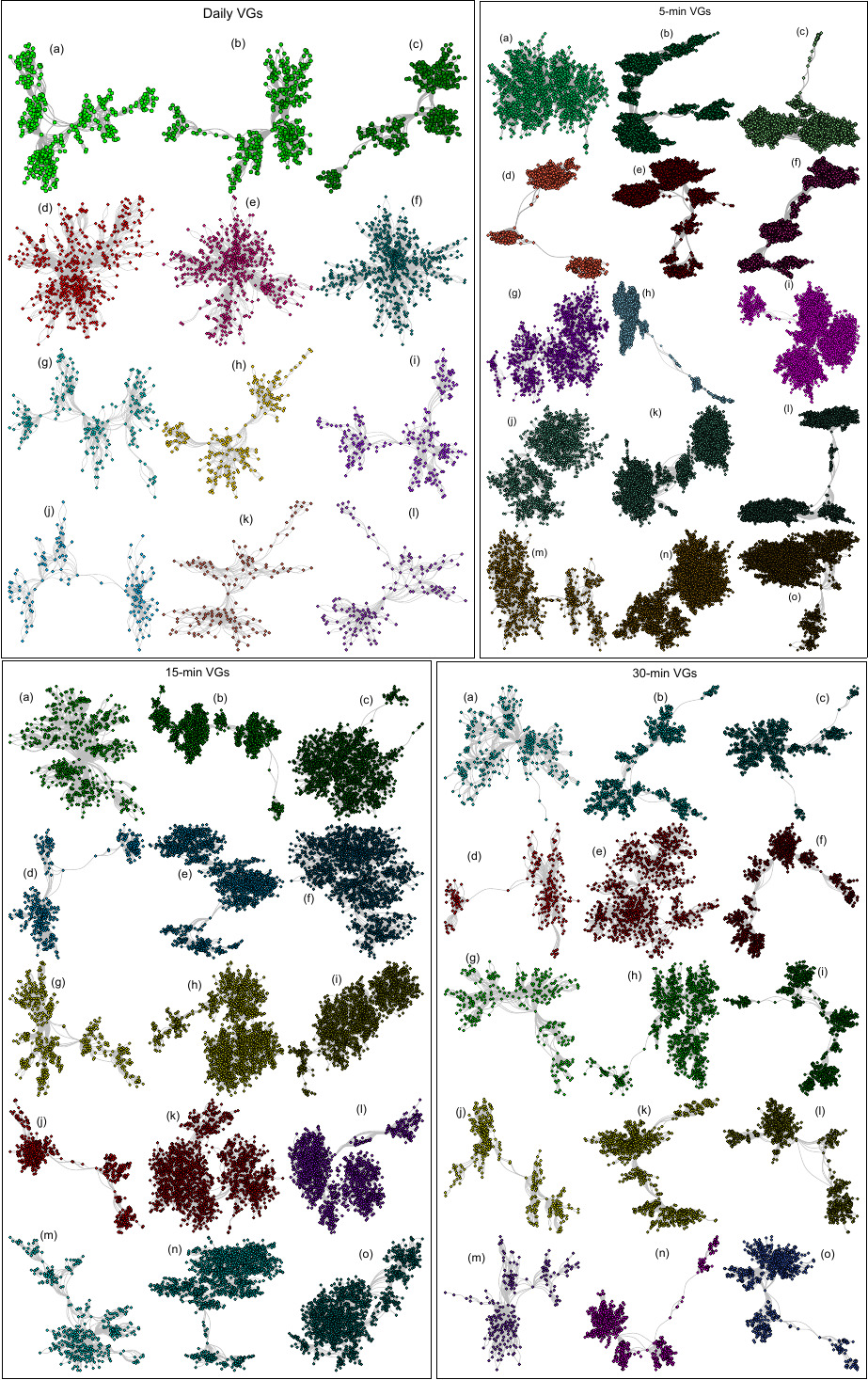}
	\caption{The VGs were constructed based on the daily, 5-minute, 15-minute, and 30-minute price data of three prominent crude oil futures: SC, WTI, and Brent. In the Daily VGs figure, panels (a-c) correspond to VGs of Sub1, panels (d-f) represent VGs of Sub2, panels (g-i) depict VGs of Sub3, and panels (j-l) illustrate VGs of Sub4. Specifically, (a), (d), (g), and (j) pertain to VGs for SC; (b), (e), (h), and (k) pertain to VGs for WTI; and (c), (f), (i), and (l) pertain to VGs for Brent. In the 5-minute VGs, 15-minute VGs, and 30-minute VGs figures, panels (a-c) correspond to VGs of period M1, panels (d-f) stand for VGs of period M2, panels (g-i) depict VGs of period M3, panels (j-l) illustrate VGs of period M4, and panels (m-o) showcase VGs of period M5. Again, (a), (d), (g), (j), and (m) represent VGs for SC; (b), (e), (h), (k), and (n) represent VGs for WTI; and (c), (f), (i), (l), and (o) represent VGs for Brent.}
	\label{Fig:VGs:Daily:Min}
\end{figure}

\section{Visibility graph construction}
\label{S3:Methodology}

As mentioned earlier, we use visibility graphs to study the fluctuation characteristics of crude oil futures prices. By utilizing visibility graphs, which bridge time series and complex networks, we can apply complex network theory to examine the global and local features of the crude oil futures price series \citep{lacasa2008time}. Next, we will describe the construction process of the visibility graph.

For a given crude oil futures price series \( P=\left\{(t_1,p_1), (t_2,p_2), \ldots, (t_i,p_i), \ldots, (t_n,p_n) \right\} \), the VG algorithm converts this price series \( P \) into an undirected complex network \( G(V,E) \). This transformation is based on the condition:
\begin{equation}
	p_k < p_j + (p_i - p_j) \frac{t_j - t_k}{t_j - t_i},
	\label{Eq:VG}
\end{equation}
where \( p_i \) denotes the price at time \( t_i \). Each time point \( t_i \) represents a node in the visibility graph \( G(V,E) \). Eq.~(\ref{Eq:VG}) provides a mathematical expression for mutual visibility between two arbitrary time points, signifying that an edge \( (t_i,t_j) \) will link nodes \( t_i \) and \( t_j \) if visibility exists between the data points \( (t_i,p_i) \) and \( (t_j,p_j) \). For a clearer understanding of the VG mapping process, Fig.~\ref{Fig:VG:Schematic} depicts a schematic representation: the upper panel shows a sequence of 16 data points, while the lower panel illustrates the corresponding VG. Fig.~\ref{Fig:VGs:Daily:Min} illustrates the VGs derived from the three crude oil futures prices at various frequencies.

\section{Empirical analysis}
\label{S4:Empirical analysis}

\subsection{Degree distributions}

Degree indicates the number of edges connected to nodes and is a key metric for understanding fundamental properties of complex networks. While node degree can be divided into in-degree and out-degree, in our analysis, the VG is an undirected graph, so we only consider its total degree, denoted as $k$.
\begin{figure}[!ht]
	\centering
	\includegraphics[width=16cm]{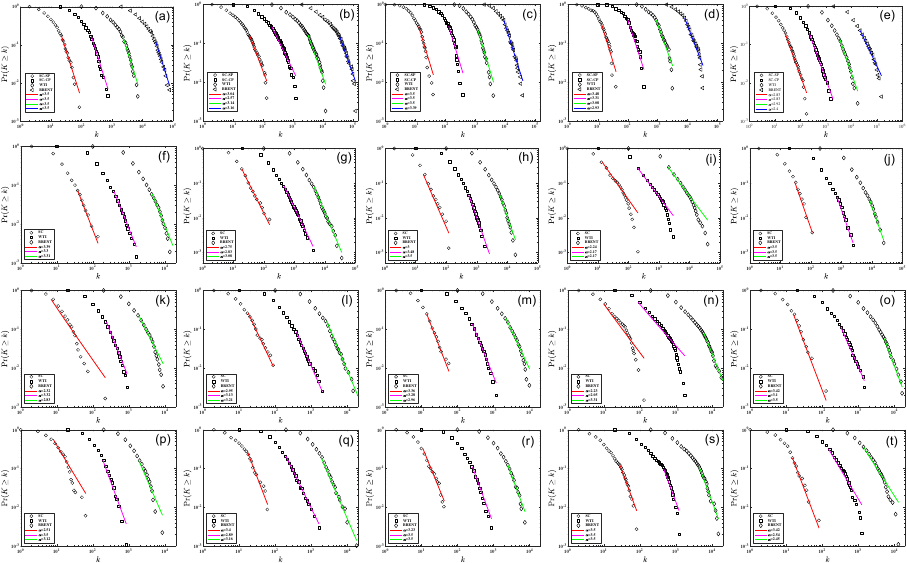}
	\caption{Empirical complementary cumulative distributions of node degree \(k\) for VGs at different frequencies, along with the fitted power law distribution. Sub-figures (a-e) correspond to daily VGs for Sub1, Sub2, Sub3, Sub4, and the entire sample, respectively. To enhance visibility, the curves for SC-CP, WTI, and Brent have been horizontally shifted by 10, 100, and 1000 times. Sub-figures (f-j) represent 5-minute VGs for M1, M2, M3, M4, and M5, with similar horizontal shifts for better visualization. Sub-figures (k-o) depict 15-minute VGs for M1, M2, M3, M4, and M5, with WTI and Brent curves shifted horizontally by 10 and 100 times. Lastly, sub-figures (p-t) illustrate 30-minute VGs for M1, M2, M3, M4, and M5, with similar horizontal shifts applied to WTI and Brent.}
	\label{Fig:VGINE:Net:Covid:Before:After:CCDF}
\end{figure}
 Assessing node degrees is essential to determine a node's centrality and influence within the network. Nodes with many connections often play critical roles in information dissemination and maintaining network stability. Thus, we begin our analysis by studying the degree distribution of the VGs for crude oil futures.

In Fig.~\ref{Fig:VGINE:Net:Covid:Before:After:CCDF}, we depict the empirical probability distributions of crude oil VGs across different frequencies. It's evident that the degree distributions of most VG networks display power-law tail characteristics, as characterized by the equation:
\begin{equation}
	p(k) = \Pr(K= k) = Ck^{-\alpha},
\end{equation}
where $\alpha$ is the scaling parameter, and $C$ is a normalization constant. After calculating the normalization constant $C$, we can express:
\begin{equation}
	p(k) = \frac{k^{-\alpha }}{\zeta(\alpha, k_{\min})},
\end{equation}
Here, $k_{\min}$ acts as the lower boundary for the scaling range governing the power-law decay, and $\zeta(\alpha, k_{\min})$ denotes the Hurwitz zeta function \citep{clauset2009power,yang2017statistical,yang2019non}. In many contexts \citep{clauset2009power}, considering the complementary cumulative distribution function (CCDF) of a power-law distributed variable proves insightful:
\begin{equation}
	P(k)=\Pr(K \geq k) = \frac{\zeta(\alpha, k)}{\zeta(\alpha, k_{\min})}\sim k^{-\alpha+1},
\end{equation}
Here, the power-law exponent of the CCDF, symbolized as $\beta$, equals $\alpha - 1$.

To determine the power-law exponent $\alpha$ in the tail of the VG network degree distribution, we adopt the robust methodology proposed by \cite{clauset2009power}. It's worth noting that a defining feature of power-law distributions is the relationship between $\alpha$ and tail decay: a larger $\alpha$ signifies faster tail decay, meaning that extreme events become less likely and smaller values become more prevalent.

The scaling parameter $\alpha$ of the degree distribution, as shown in Fig.~\ref{Fig:VGINE:Net:Covid:Before:After:CCDF}, fluctuates between 2 and 3.5. It's worth noting that the power-law exponent \( \alpha \) for the three crude oil futures VGs was consistently 3.5 in the pre-COVID-19 outbreak Sub1, while in the post-COVID-19 outbreak Sub2, all values of \( \alpha \) decreased. Fig.~\ref{Fig:VGINE:Net:Covid:Before:After:CCDF} (c) corresponds to the interval following the Russia-Ukraine conflict outbreak. When compared to Sub1, the scaling exponent \( \alpha \) did not decrease significantly. The interval depicted in Fig.~\ref{Fig:VGINE:Net:Covid:Before:After:CCDF} (d) represents the period of overlap between the regular prevention of COVID-19 and the Russia-Ukraine conflict. Compared to Sub1, it's noticeable that the scaling exponent \( \alpha \) has marginally decreased. Specifically, when we observe Fig.~\ref{Fig:VGINE:Net:Covid:Before:After:CCDF} (e), it becomes evident that the tail decay of the degree distribution has slowed down, with all power-law exponents (\( \alpha \)) being less than 3. This suggests that during the entire sample period, relatively frequent occurrences of extreme prices can be discerned, with these extreme prices possibly influenced by major events like the COVID-19 pandemic and the Russia-Ukraine conflict.

Additionally, an assessment of the scaling parameter \( \alpha \) in the VG degree distribution at 5-minute, 15-minute, and 30-minute intervals reveals that the advent of the COVID-19 pandemic has undeniably augmented the incidence rate of extreme price fluctuations across all crude oil futures markets. This observation is further supported by a close examination of the values of \( \alpha \) in Fig.~\ref{Fig:VGINE:Net:Covid:Before:After:CCDF} (f-g), (k-l), and (p-q), where many subplots clearly display power-law characteristics in their tail distributions.

\subsection{Clustering coefficient}

\begin{figure}[!ht]
		\centering
		\includegraphics[width=16cm]{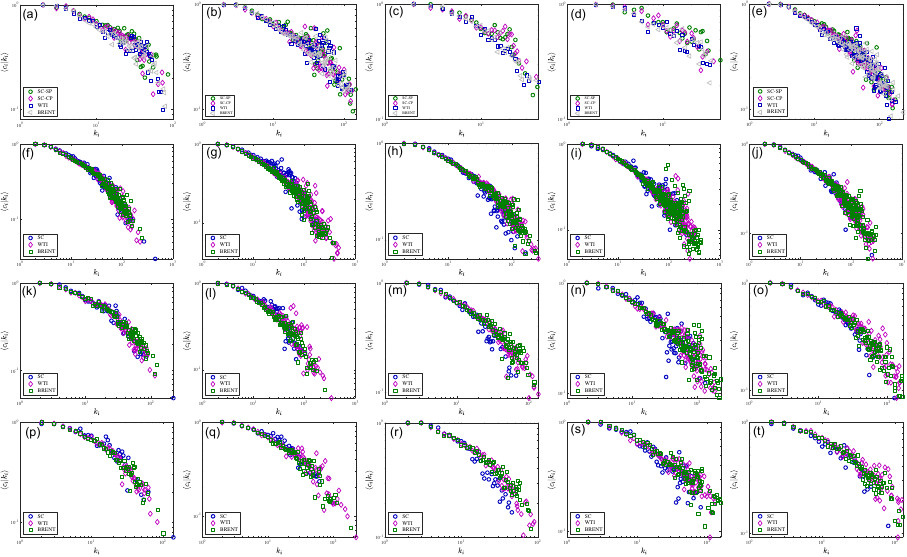}
		\caption{ Relationship between the node clustering coefficient $\langle c_i|k_i \rangle$ and node degree $k_i$ for VGs across different sampling frequencies. Panels (a-e) display daily VGs from Sub1 through Sub4, concluding with the entire sample. Panels (f-j) stand for 5-minute VGs of M1, M2, M3, M4, and M5. Panels (k-o) correspond to 15-minute VGs of M1, M2, M3, M4, and M5. Panels (p-t) are 30-minute VGs of M1, M2, M3, M4, and M5. }
		\label{Fig:VG:k:c}
\end{figure}

The node clustering coefficient $c_i$ is a metric used to quantify the likelihood of interconnections among the immediate neighbors of an individual node within a network. This metric has been extensively explored in various network science studies, contributing to our understanding of network structures \citep{watts1998collective, newman2003structure}. It is computed using the following formula:
\begin{equation}
	c_i= \frac{2 \cdot E_i}{k_i \cdot (k_i-1)}
\end{equation}
where \(E_i\) denotes the count of existing edges among the neighbors of node \(i\), and \(k_i\) stands for the degree of node \(i\). 

The global clustering coefficient is another essential metric, especially when analyzing the clustering properties of networks. This coefficient provides an assessment of the overall clustering tendency or triadic closure in an entire network. It can be determined by calculating the average of the node clustering coefficients $c_i$ for all nodes:
\begin{equation}
	C =\langle c_i \rangle= \frac{1}{N} \sum_{i=1}^{N} c_i
\end{equation}
where \(N\) denotes the total number of nodes in the network. The reported coefficients $C$ in Table~\ref{TB:VG:ClusteringCoefficient} show that the global clustering coefficients across various frequencies and sample intervals consistently hover around 0.7, peaking at 0.7442. This observation implies an increased prevalence of triangular connections among VG network nodes, suggesting that nodes in the VGs typically form tightly-knit clusters or communities. This interpretation is supported by the graphical representations of VGs in Fig.~\ref{Fig:VGs:Daily:Min}.

	\begin{table}[!htp]\addtolength{\tabcolsep}{2pt}
	\footnotesize
	\caption{Global clustering coefficient $C$ and assortativity coefficient $r$.}
	\label{TB:VG:ClusteringCoefficient}
	\resizebox{\textwidth}{!}{
		\medskip
		\centering
		\begin{threeparttable}
			\begin{tabular}{cccccccccccccccccccccccccccccc}
				\hline\hline
				\multirow{3}*[2mm]{Name}&\multicolumn{5}{c}{SC-SP}&&\multicolumn{5}{c}{SC-CP}&&\multicolumn{5}{c}{WTI}&&\multicolumn{5}{c}{Brent}\\
				\cline{2-6}\cline{8-12}\cline{14-18}\cline{20-24}
				&Sub1 & Sub2 & Sub3 & Sub4 & Whole && Sub1 & Sub2 & Sub3 & Sub4 & Whole && Sub1 & Sub2 & Sub3 & Sub4 & Whole && Sub1 & Sub2 & Sub3 & Sub4 & Whole  \\
				\hline
				\multicolumn{16}{l}{\textit{Panel A : Daily data}} \\
				$C$&0.7229& 0.6649& 0.7203 & 0.7152 & 0.6873 && 0.7282& 0.6776& 0.7205 & 0.7306 & 0.7001&& 0.7020& 0.6797& 0.7352 & 0.7308 & 0.6898&& 0.7005& 0.6747& 0.7442 & 0.7291 & 0.1799\\
				$r$&0.2473& 0.0205& 0.1208 & 0.1746 & 0.1106 && 0.1475& 0.0611& 0.1645 & 0.1190& 0.1103&& 0.2632& 0.1219& 0.1062 & 0.1753& 0.1782&& 0.1657& 0.1080& 0.0762 & 0.1642 & 0.6873\\\\			
				
				\multirow{3}*[2mm]{}&\multicolumn{5}{c}{SC}&&\multicolumn{5}{c}{WTI}&&\multicolumn{5}{c}{Brent}\\
				\cline{2-6}\cline{8-12}\cline{14-18}
				&M1 & M2 & M3 & M4 & M5&& M1 & M2 & M3 & M4 & M5&& M1 & M2 & M3 & M4 & M5 \\\\
				\multicolumn{4}{l}{\textit{Panel B: 5-minute data}} \\\
				$C$&0.7254& 0.7326& 0.7308 & 0.7013 & 0.7148 && 0.7221& 0.6994 & 0.7005 & 0.6893& 0.6998&& 0.7147 & 0.7041 & 0.7007& 0.6854& 0.6991 \\
				$r$&-0.0615& -0.1184& 0.0675 & 0.2041 & 0.0162 && 0.0441& 0.0311 & 0.1421 & 0.3353& 0.2476&& 0.1470 & 0.0853 & 0.2080& 0.1737& 0.2416 \\
				\\
				\multicolumn{4}{l}{\textit{Panel C: 15-minute data}} \\\
				
				$C$&0.7346& 0.7322& 0.7378 & 0.7041 & 0.7289 && 0.7242& 0.7069 & 0.7121 & 0.6928& 0.7110&& 0.7230 & 0.7101 & 0.7135& 0.6854& 0.7040 \\
				$r$&-0.0628& -0.0814& 0.0727 & 0.2135 & -0.0055 && 0.0664&-0.0078 & 0.1202 & 0.2713& 0.2144&& 0.1928 & 0.0591 & 0.2054& 0.1889& 0.2120 \\
				\\
				\multicolumn{4}{l}{\textit{Panel D: 30-minute data}} \\\
				
				$C$&0.7504& 0.7404& 0.7432 & 0.7086 & 0.7242 && 0.7319& 0.7173 & 0.7225 & 0.6995& 0.7140&& 0.7257 & 0.7253 & 0.7193& 0.6998& 0.7083 \\
				$r$&-0.0763& -0.0687& 0.0794& 0.1518 &0.0096 && 0.0612&-0.0131 & 0.0939 & 0.2924& 0.2393&& 0.1523 & 0.0371 & 0.1514& 0.1859& 0.1958 \\
				\hline\hline
			\end{tabular}
			\begin{tablenotes}[para,flushleft]
			\end{tablenotes}
	\end{threeparttable}}
\end{table}

Delving deeper into the local topological features of VG networks, Fig.~\ref{Fig:VG:k:c} showcases the relationship between node degree $k_i$ and the clustering coefficient $\langle c_i|k_i \rangle$ in the double-logarithmic coordinate system. Here, $\langle c_i|k_i \rangle$ represents the conditional mean of the node clustering coefficient $c_i$ given the node degree $k_i$. A clear negative relationship between $k_i$ and $\langle c_i|k_i \rangle$ is evident across all VG frequencies, indicating that nodes with smaller degrees are more likely to form denser clusters \citep{liu2023visibility}. This inverse relationship becomes even more pronounced with larger values of $k_i$. The effects of external factors, such as the COVID-19 pandemic and the Russia-Ukraine conflict, on the relationship between node degree and clustering coefficient ($k_i$ vs. $\langle c_i|k_i \rangle$) seem negligible.

\begin{figure}[!ht]
	\centering
	\includegraphics[width=16cm]{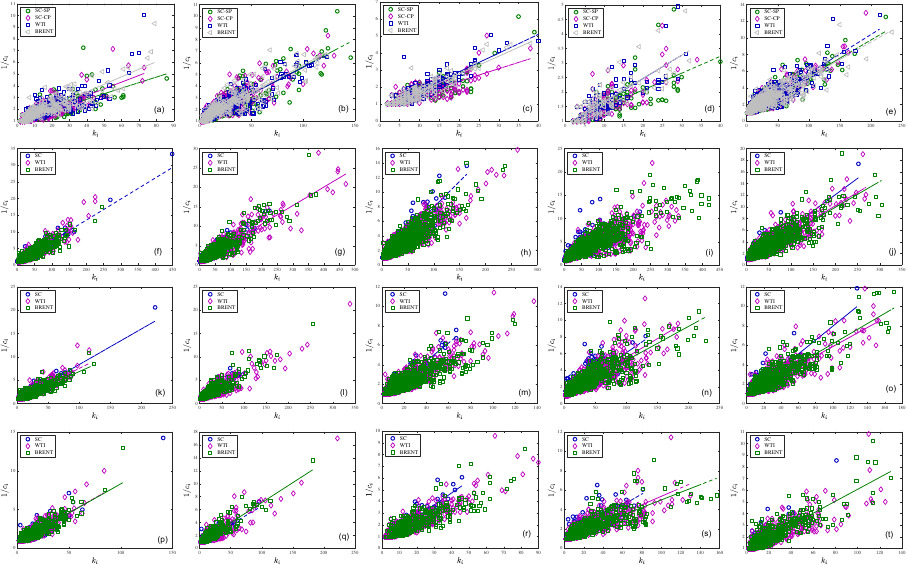}
	\caption{The relationship between the reciprocal of the clustering coefficient $1/c_i$ and the node degree $k_i$ is depicted for VGs across various frequencies. Panels (a-e) correspond to daily VGs for sub-sample 1, sub-sample 2, sub-sample 3, sub-sample 4, and the whole sample, respectively. Panels (f-j) represent 5-minute VGs for M1, M2, M3, M4, and M5. Panels (k-o) depict 15-minute VGs for M1, M2, M3, M4, and M5. Finally, panels (p-t) showcase 30-minute VGs for M1, M2, M3, M4, and M5. }
	\label{Fig:VG:k:reciprocal}
\end{figure}

Furthermore, Fig.~\ref{Fig:VG:k:reciprocal} aligns with Fig.~\ref{Fig:VG:k:c}, showing a consistent positive linear relationship between node degree $k_i$ and the reciprocal of the clustering coefficient $1/c_i$. This relationship suggests nodes have a propensity to form densely interconnected local cliques. Specifically, for the daily frequency samples, this positive relationship exists in at least two of the three crude oil futures price VGs. However, in the high-frequency 5-minute, 15-minute, and 30-minute VGs, at least one VG demonstrates this linear relationship, with the exceptions being sub-figures (i) and (l) in Fig.~\ref{Fig:VG:k:reciprocal}.

\subsection{Small-world property}

\begin{figure}[!htp]
	\centering
	\includegraphics[width=14cm]{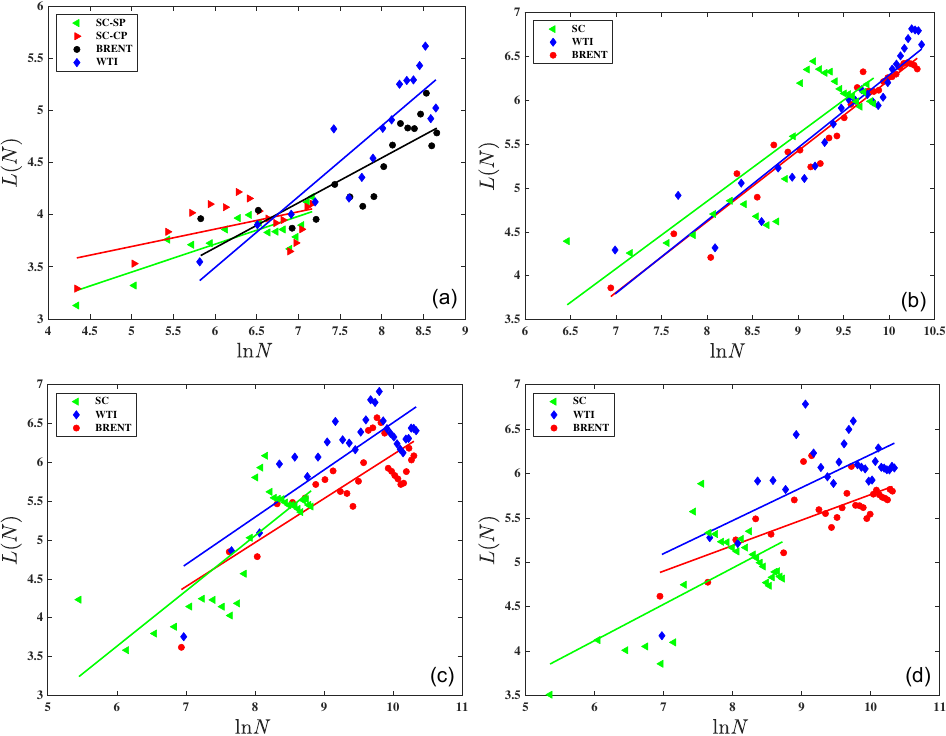}
	\caption{Dependence of the average shortest path length $L(N)$ on the total number of nodes $N$ of VGs at different frequencies. (a) Daily VGs. (b) 5-minute VGs. (c) 15-minute VGs. (d) 30-minute VGs.}
	\label{Fig:VGINE:Net:Covid:Before:After:K:averageC}
\end{figure}

The small-world property was first introduced by \cite{watts1998collective}, emphasizing the balance between localized connections and efficient global connectivity within networks. A VG network exhibiting the small-world property typically showcases two primary characteristics. Firstly, the VG's clustering coefficients $C$ are significantly elevated, as presented in Table~\ref{TB:VG:ClusteringCoefficient}. Secondly, the average shortest path length $L(N)$ for the VG should satisfy the relationship
\begin{equation}
	L(N) \propto \ln N,
\end{equation}
where $N$ represents the total number of nodes in the VG network. The formula for computing  $L(N)$ is given by
\begin{equation}
	L(N) = \frac{1}{N(N-1)} \sum_{i \neq j} d(i,j),
\end{equation}
where $d(i,j)$ represents the shortest path length between any two distinct nodes ($i$ and $j$, with $i$ not equal to $j$) within the VG network. Fig.~\ref{Fig:VGINE:Net:Covid:Before:After:K:averageC} portrays the linear relationship between the average shortest path length $L(N)$ and the logarithm of the number of nodes $\ln N$ across diverse frequencies. A strong positive linear correlation between $\ln N$ and $L(N)$ is evident for each VG in Fig.~\ref{Fig:VGINE:Net:Covid:Before:After:K:averageC}, solidifying the claim that VG networks across all frequencies adhere to the small-world property.

\subsection{Mixing pattern}

\begin{figure}[!htb]
	\centering
	\includegraphics[width=16cm]{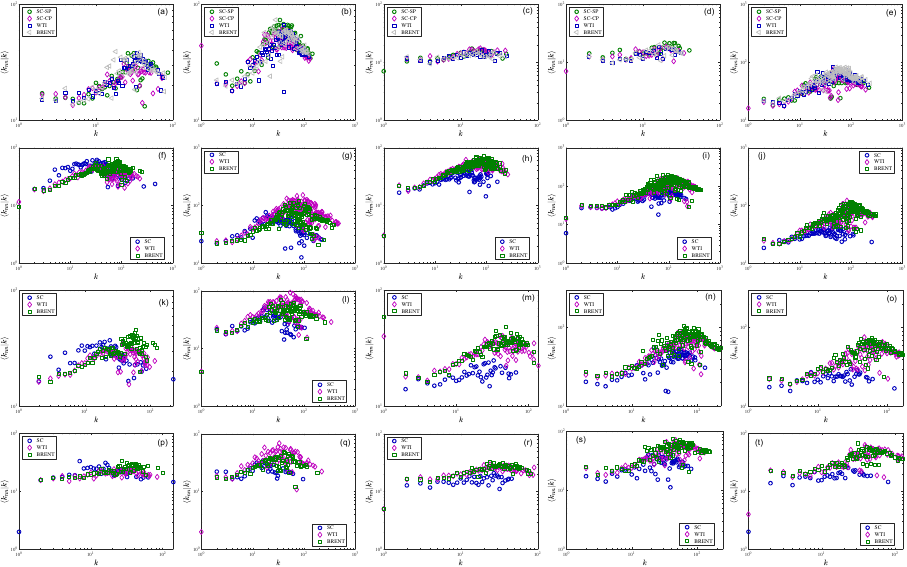}
	\caption{The variation of the average nearest neighbor degree $\langle k_{nn}|k \rangle$ as a function of node degree $k$ is presented, showcasing VGs across various frequency regimes. Panels (a-e) correspond to the daily VGs of sub-samples 1, 2, 3, 4, and the entire sample, respectively. Panels (f-j) showcase 5-minute VGs for M1, M2, M3, M4, and M5. Panels (k-o) exhibit 15-minute VGs for M1, M2, M3, M4, and M5, while panels (p-t) elucidate the 30-minute VGs for M1, M2, M3, M4, and M5. }
	\label{Fig:VG:Mix:Pattern}
\end{figure}

Understanding assortative mixing patterns is essential in network analysis, as it can expose critical structural properties and dynamics within networks, such as community formation, information flow, and the spread of influence. An assortative mixing pattern in a network indicates the propensity of nodes to associate with others that possess similar characteristics or attributes. In assortative networks, nodes with similar properties have a higher likelihood of being linked, while in disassortative networks, nodes with distinct properties tend to connect. Measures like the assortativity coefficient or assortativity degree can quantify assortative mixing patterns \citep{newman2002assortative}, offering insights into node organization based on their properties.

To elucidate the mixing pattern inherent in VGs across various frequencies, we employ the average nearest-neighbor degree, symbolized as $\langle k_{nn}|k \rangle$. To compute $\langle k_{nn}|k \rangle$, one must first determine the average degree of adjacent nodes $k_{nn,i|k}$ related to node $i$. The mathematical representation for $k_{nn,i|k}$ is
\begin{equation}
	k_{nn,i|k}  = \frac{1}{k_i} \sum_{j \in N_i} k_j,
\end{equation}
with $N_i$ representing the set of neighboring nodes of node $i$ and $j$ being a neighbor of $i$. The formula for the average nearest neighbor degree for nodes with a specified degree $k$ is
\begin{equation}
	\langle k_{nn}|k \rangle= \frac{1}{N_{k_i}} \sum_{k_i=k} k_{nn,i},
\end{equation}
with \(N_{k_i}\) signifying the number of nodes with a node degree of $i$. Assortative mixing is indicated if $\langle k_{nn|k} \rangle$ rises with $k$, while a decline suggests disassortative mixing.

Fig.~\ref{Fig:VG:Mix:Pattern} delineates the evolution of $\langle k_{nn}|k \rangle$ as a function of node degree $k$ for VGs across distinct frequency domains. Each subplot in Fig.~\ref{Fig:VG:Mix:Pattern} reveals a common trend, irrespective of the VG's frequency regime. $\langle k_{nn}|k \rangle$ undergoes a steady decline with an increasing node degree $k$ before rising to a particular value of ${\langle k_{nn}|k \rangle}_{\rm critical}$. Then, it decreases once more, revealing intricate assortative mixing.

The multifaceted assortativity patterns discerned across various VGs underscore the differences in the repercussions of both the COVID-19 pandemic and the Russia-Ukraine conflict on the trio of crude oil markets. To further differentiate these effects, we employ Newman's assortativity coefficient, depicted as 
\begin{equation}
	r = \frac{{M^{-1}\sum_{i} j_ik_i  - \left[  M^{-1}\sum_i \frac{1}{2}(j_i + k_i)\right]^2}}{{M^{-1}\sum_i \frac{1}{2}(j_i^2 + k_i^2) -\left[  M^{-1}\sum_i \frac{1}{2}(j_i + k_i)\right]^2 }},
\end{equation}
as an additional descriptor \citep{newman2002assortative}. Here, $r$ denotes the Pearson correlation coefficient between degrees and ranges from $-1$ to $1$. The terms $j_i$ and $k_i$ pertain to the degrees of the nodes at the endpoints of the $i$th edge, with $i$ extending from 1 to $M$.

Data from Table~\ref{TB:VG:ClusteringCoefficient} unveils that during Sub2, the daily assortativity coefficients $r$ for SC-SP and SC-CP are exceptionally low at 0.0205 and 0.0611, respectively, hovering near zero. These metrics underline the profound influence of the COVID-19 pandemic on the Shanghai crude oil market, resulting in virtually nonexistent assortative mixing. In contrast, the assortativity coefficients $r$ for Sub3 hint at a milder impact of the Russia-Ukraine crisis on the SC. Notably, a focused view on the daily assortativity coefficients $r$ for WTI and Brent reveals that Sub3 possesses the lowest $r$ values, while Sub2 presents intermediate figures. This observation signifies that the combined ramifications of the Russia-Ukraine conflict and the COVID-19 pandemic more starkly affect WTI and Brent compared to the singular impact of the pandemic. Analyzing these events' implications on the trio of crude oil markets further illuminates the disparities between the emergent crude oil futures market (SC) and the established mature markets (WTI and Brent) in the long run. However, the calculated assortativity coefficient $r$ based on 5-minute, 15-minute, and 30-minute data from short-term M2 and M4, as depicted in Table~\ref{TB:VG:ClusteringCoefficient}, suggests that both events similarly impacted the trio of crude oil futures markets in the short term. This alignment indicates that these events led to a greater propensity for price data points to link with others displaying analogous pricing behaviors.

\begin{figure}[!htb]
	\centering
	\includegraphics[width=14.2cm]{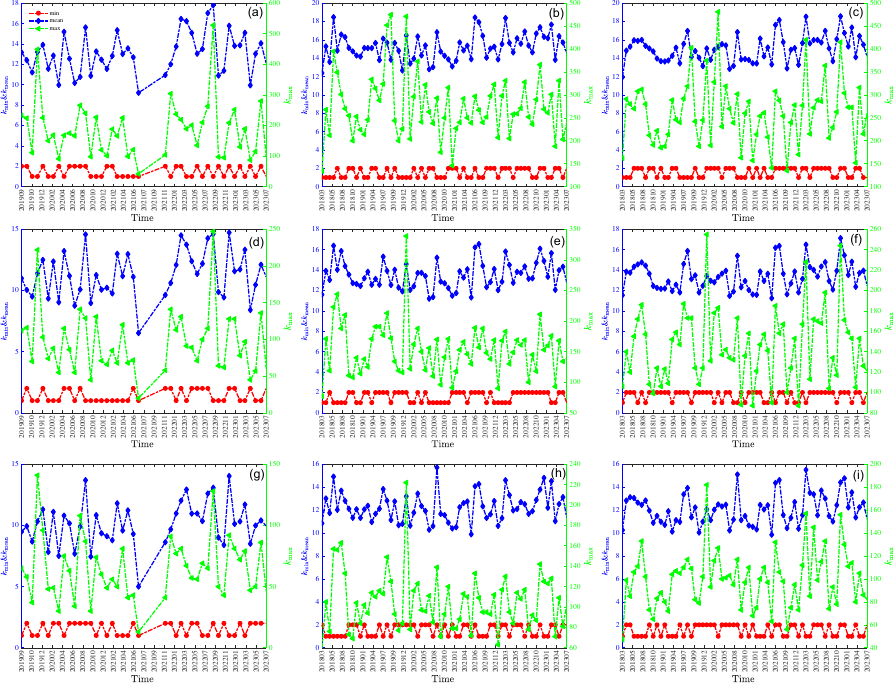}
	\caption{The evolution of node degrees $k_i$ is observed within monthly time windows. Subfigures (a-c) correspond to 5-minute VGs, (d-f) to 15-minute VGs, and (g-i) to 30-minute VGs. Subfigures (a), (d), and (g) represent VGs related to SC, while (b), (e), and (h) pertain to WTI, and (c), (f), and (i) are associated with Brent. }
	\label{Fig:VG:degree:evolution}
\end{figure}

\subsection{Time-varying VGs}

With the high-frequency sample data at our disposal, we delve into the time-varying characteristics of VG networks, transcending the static realm. Fig.~\ref{Fig:VG:degree:evolution} illustrates the temporal evolution of node degrees at monthly intervals for all high-frequency VGs. We've used dual axes to distinguish between $k_{\rm min}$, $k_{\rm mean}$, and $k_{\rm max}$. The monthly minimum degrees $k_{\rm min}$ fluctuate between 0 and 1 for all VGs. Average degrees for 5-minute, 15-minute, and 30-minute VGs hover around 15, 14, and 12, respectively. The maximum degrees show varying fluctuations based on VG frequency. As data frequency increases, both $k_{\rm mean}$ and $k_{\rm max}$ also increase due to the surge in data points and connections. The unique trading time mechanism for SC results in fewer minute-level data points within a day, yielding lower overall node degrees. While VGs in all markets display similar evolutionary patterns, SC differs slightly from WTI or Brent. For instance, the spike in maximum degrees for WTI and Brent VGs in January 2020 highlights the pandemic's profound influence on crude oil futures markets, a phenomenon absent in the SC market. Notably, only Brent's maximum degree surged in March 2022, marking the Russia-Ukraine conflict's imprint.

\begin{figure}[h]
	\centering
	\includegraphics[width=14.2cm]{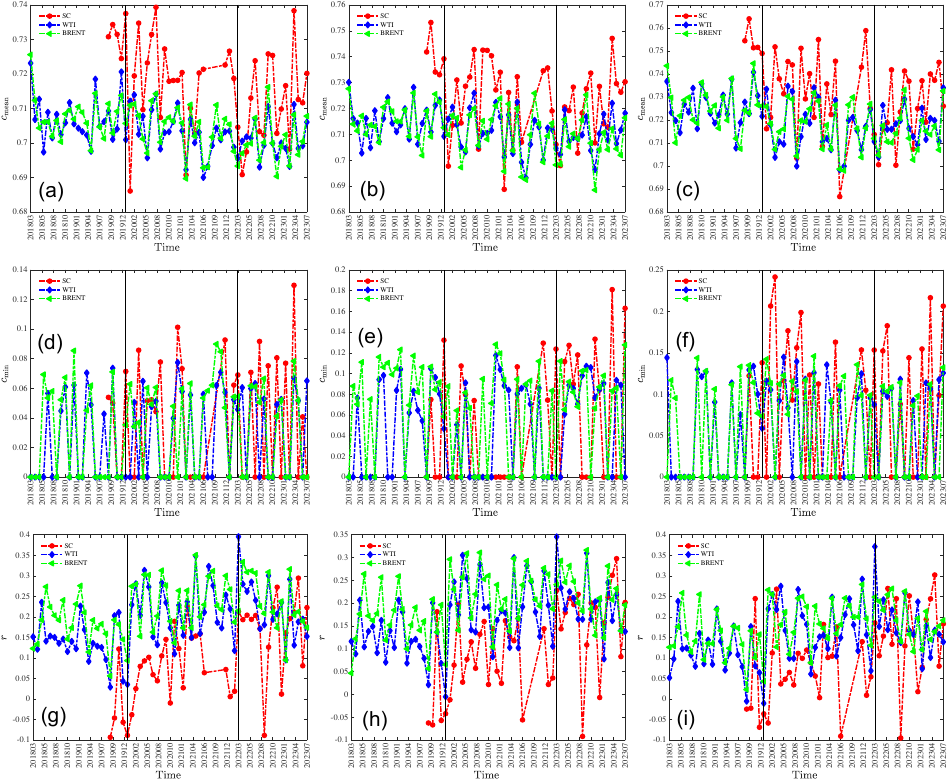}
	\caption{Temporal variations in clustering coefficients $c_i$ and assortativity coefficients $r$ are depicted at monthly intervals. Subfigures (a), (b), and (c) portray the mean clustering coefficients $c_{\rm mean}$ for 5-minute, 15-minute, and 30-minute VGs, while subfigures (d), (e), and (f) illustrate the minimum clustering coefficients $c_{\rm min}$ of the corresponding VGs. Subfigures (g), (h), and (i) present assortativity measurements based on 5-minute, 15-minute, and 30-minute VGs, respectively. In addition, the two vertical lines in each subgraph correspond to the most recent month after the outbreak of the COVID-19 pandemic and the outbreak of the Russia-Ukraine conflict, in that order.}		
	\label{Fig:VG:clustering:assort:evolution}
\end{figure}

In Fig.~\ref{Fig:VG:clustering:assort:evolution}, we observe monthly fluctuations in clustering coefficients $c_i$ and assortativity coefficients $r$. The maximum clustering coefficients $c_{\rm max}$ for all VG frequencies stand at 1, and are hence omitted for succinctness. Monthly VGs of various frequencies consistently display a high degree of clustering, a finding corroborated by Fig. \ref{Fig:VGs:Daily:Min}. Both WTI and Brent show similar trajectories for $c_{\rm mean}$ and $c_{\rm min}$, but SC often posts higher $c_{\rm mean}$ values. As illustrated in Fig. \ref{Fig:VG:k:reciprocal} and Fig. \ref{Fig:VG:k:c}, a negative correlation exists between $c_i$ and $k_i$. Thus, SC's higher $c_{\rm mean}$ values in Fig.\ref{Fig:VG:clustering:assort:evolution} align with the lower $k_{\rm max}$ and $k_{\rm mean}$ values observed in Fig.~\ref{Fig:VG:degree:evolution}. The notable jumps in SC's $c_{\rm mean}$ in December 2019 and January 2020 emphasize the COVID-19 pandemic's impact.

Continuing with assortativity coefficient $r$, similar patterns to $c_{\rm mean}$ evolution emerge. Both WTI and Brent show almost identical trends, while SC remains consistently lower. Monthly VGs of various frequencies revolve around an average of 0.2, indicating subtle assortative tendencies in high-frequency VGs. Notably, the 5-minute VGs vividly capture the significant effects of both the COVID-19 pandemic and the Russia-Ukraine conflict on the three crude oil futures markets.

\section{Conclusion}
\label{S5:conclusion}

In this detailed analysis of the three crucial crude oil futures markets, WTI, Brent, and SC, we employed the visibility graph methodology to gain deep insights. Through both daily and high-frequency samples, we delved into the static and dynamic properties of these markets, uncovering significant patterns in their structure and evolution.

Examining the static characteristics, we meticulously analyzed degree distributions, clustering coefficients, assortativity coefficients, and small-world properties. Most VGs demonstrated a clear power-law decay in their degree distributions, with the scaling parameter $\alpha$ varying between 2 and 3.5. We recognized the pronounced clustering tendencies in crude oil futures VGs and verified the inverse relationship between the clustering coefficient and node degree. Furthermore, the assortativity coefficient $r$ and the average nearest neighbor degree $\langle k_{nn}|k \rangle$ together indicated that VGs, irrespective of their frequency, display complex assortative mixing patterns. The marked small-world properties across all markets highlight their capability for swift information transfer. These static properties also echo the distinct influences of both the COVID-19 pandemic and the Russia-Ukraine conflict on the markets.

Turning to dynamic features, we observed captivating shifts in degree distributions, clustering coefficients, and assortativity coefficients over time. The closely mirroring patterns between WTI and Brent suggest their behavioral alignment, while SC's unique dynamics can be attributed to its distinct trading procedures. Specifically, the 5-minute VGs' assortativity coefficient $r$ stands out in reflecting the major impacts of the COVID-19 pandemic and the Russia-Ukraine conflict on the crude oil futures markets. Within this context, the assortativity coefficient $r$ seems to resonate with the short-term market responses to these events. Nevertheless, a holistic grasp of these events' influence on market volatility would require an expanded study considering other factors like price shifts, trading duration, volumes, and more. During the Russia-Ukraine conflict, discernible variations in network properties were evident across the markets, underscoring the differing sensitivities of these markets to geopolitical shifts.

Our research illuminates the multifaceted nature of these markets, delineating their distinctive features and reactions to external stimuli. This study enriches the domain of finance by elucidating the structural and temporal intricacies of these vital commodities markets, emphasizing their adaptability amidst global adversities.
For future perspective research, expanding the visibility graph methodology to other crude oil futures markets to verify its universality would be beneficial. Simultaneously, delving into the relationship between extreme events and network properties using advanced statistical or machine learning methods will ensure a deeper understanding of market dynamics in the evolving financial landscape.

\section*{Acknowledgements}

This work was supported by the Shanghai Planning Office of Philosophy and Social Science (2022EJB006).

\end{document}